\documentclass[twocolumn,english,aps,prl,showpacs,superscriptaddress,nobibnotes,longbibliography]{revtex4-1}
\usepackage[T1]{fontenc}
\usepackage[latin9]{inputenc}
\usepackage{float}
\usepackage{units}
\usepackage{amsmath}
\usepackage{amssymb}
\usepackage{graphicx}

\makeatletter
\usepackage{babel}

\makeatother

\usepackage{babel}
\begin{document}
\title{Atom interferometry with thousand-fold increase in dynamic range}
\author{Dimitry Yankelev}
\thanks{These authors contributed equally to this work.~\\
dimitry.yankelev@weizmann.ac.il~\\
chen.avinadav@weizmann.ac.il}
\affiliation{Department of Physics of Complex Systems, Weizmann Institute of Science,
Rehovot 7610001, Israel}
\affiliation{Rafael Ltd, Haifa 3102102, Israel}
\author{Chen Avinadav}
\thanks{These authors contributed equally to this work.~\\
dimitry.yankelev@weizmann.ac.il~\\
chen.avinadav@weizmann.ac.il}
\affiliation{Department of Physics of Complex Systems, Weizmann Institute of Science,
Rehovot 7610001, Israel}
\affiliation{Rafael Ltd, Haifa 3102102, Israel}
\author{Nir Davidson}
\affiliation{Department of Physics of Complex Systems, Weizmann Institute of Science,
Rehovot 7610001, Israel}
\author{Ofer Firstenberg}
\affiliation{Department of Physics of Complex Systems, Weizmann Institute of Science,
Rehovot 7610001, Israel}
\begin{abstract}
The periodicity inherent to any interferometric signal entails a fundamental
trade-off between sensitivity and dynamic range of interferometry-based
sensors. Here we develop a methodology for significantly extending
the dynamic range of such sensors without compromising their sensitivity,
scale-factor, and bandwidth. The scheme is based on operating two
simultaneous, nearly-overlapping interferometers, with full-quadrature
phase detection and with different but close scale factors. The two
interferometers provide a joint period much larger than $2\pi$ in
a moiré-like effect, while benefiting from close-to-maximal sensitivity
and from suppression of common-mode noise. The methodology is highly
suited to atom interferometers, which offer record sensitivities in
measuring gravito-inertial forces but suffer from limited dynamic
range. We experimentally demonstrate an atom interferometer with a
dynamic-range enhancement of over an order of magnitude in a single
shot and over three orders of magnitude within a few shots, for both
static and dynamic signals. This approach can dramatically improve
the operation of interferometric sensors in challenging, uncertain,
or rapidly varying, conditions.
\end{abstract}
\maketitle
The ambiguity-free dynamic range of interferometric physical sensors
is fundamentally limited to $2\pi$ radians. When the \emph{a priori}
phase uncertainty is larger than a single fringe, additional information
is required to uniquely determine the physical quantity measured by
the interferometer. If this quantity remains constant over long periods
of time, the phase ambiguity may be resolved through additional interferometric
measurements with different scale-factors, defined as the ratio between
the interferometer phase and the magnitude of the physical quantity.
A more challenging scenario arises when the physical quantity changes
rapidly with time, and measurement with multiple scale-factors must
be realized simultaneously.

Overcoming this challenge in cold-atom interferometers \citep{Varenna2014},
which have emerged over the past decades as extremely sensitive sensors
of gravitational and inertial forces, is an especially ambitious proposition.
Applications of atom interferometers vary from fundamental research
\citep{Mueller2010,Kovachy2015,Barrett2016,Becker2018,Xu2019} and
precision measurements \citep{Rosi2014,Parker2018} to gravity surveys
and inertial navigation \citep{Geiger2020}. Mobile interferometers
are being developed by several groups \citep{Bongs2019,Farah2014,Freier2016,Menoret2018}
with demonstrations of land-based, marine, and airborne gravity surveys
\citep{Bidel2018,Wu2019,Bidel2020}.

In these applications, limited dynamic range is especially challenging,
as the uncertainty in the acceleration to be measured is potentially
very large. Reducing the interferometer scale-factor or performing
multiple measurements at each location results in reduced sensitivity
or lower temporal bandwidth, respectively. A common solution relies
on auxiliary sensors with larger dynamic range but lower resolution
to constrain the interferometric measurement to a smaller, non-ambiguous
range \citep{Merlet2009,Lautier2014}. However, this approach may
suffer from transfer-function errors, misalignment between the sensors,
or non-linearities \citep{Bidel2018}. It is therefore highly desirable
to have a high-sensitivity, high-bandwidth, atom interferometer with
a large dynamic range. While optical interferometers may gain such
capabilities by employing and detecting multiple wavelengths \citep{Daendliker1999,Falaggis2009},
this feat is more challenging for matter-wave interferometers.

\begin{figure*}[t]
\begin{centering}
\includegraphics[width=2\columnwidth]{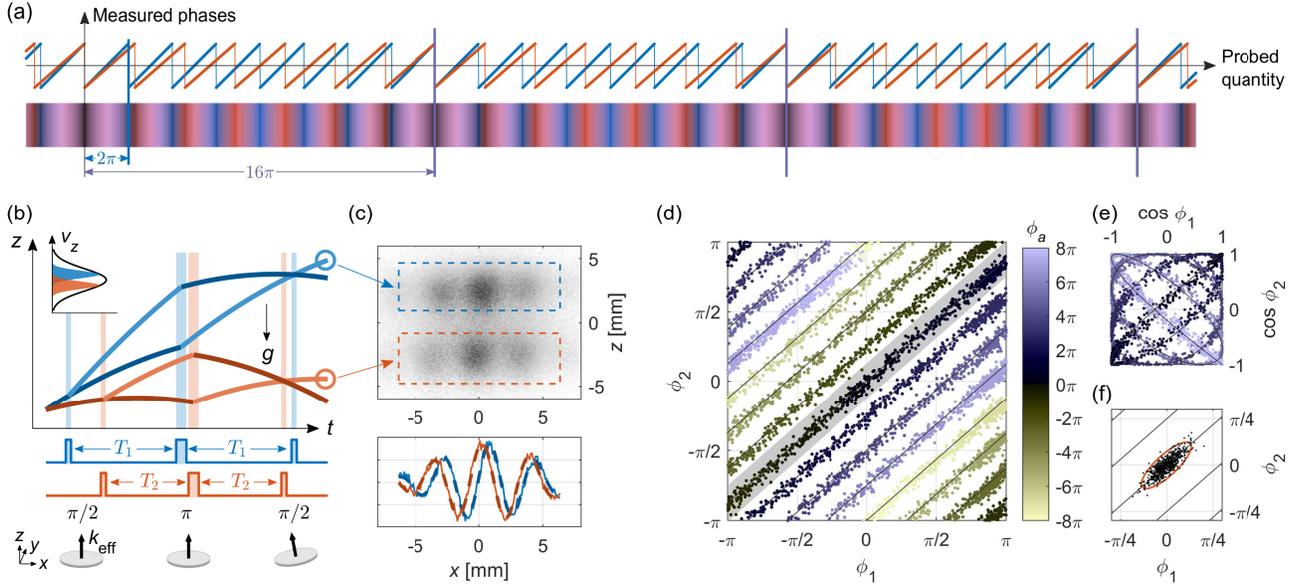}
\par\end{centering}
\centering{}\caption{Concept, scheme and results of dual-$T$ interferometry. (a) Conceptual
representation of dynamic-range enhancement by a factor of $\times8$,
using a pair of simultaneous interferometers with different scale
factors. (b) Dual-$T$ atom interferometry. A pair of Raman pulse
sequences (red and blue), with different interrogation times $T$
and addressing different velocity classes of the atoms, couple between
two atomic states with momentum difference of $\hbar k_{\textrm{eff}}$
(bright and dark trajectories). To obtain full phase quadrature information,
the Raman retro-reflecting mirror is tilted before the final $\pi/2$
pulses, generating a transverse phase gradient across the cloud. (c)
Top: a single fluorescence image captures the population in one of
the atomic states for both interferometers. Bottom: measured cross-sections
(solid lines) and the fitted fringes (dashed lines) of both interferometers,
after vertical integration of the regions indicated by the dashed
rectangles and subtraction of the Gaussian envelope. The interferometer
phases $\phi_{1},\phi_{2}$ are determined by the fringe phase at
the center of each cloud. (d) Results of dual-$T$ interferometry
for inertial phase $\phi_{a}$ in the range of $\pm8\pi$ (color coded),
each dot represents a single dual-$T$ measurement. Slope of gray
lines is the scale-factor ratio $\tau=\left(T_{2}/T_{1}\right)^{2}=7/8$.
Shaded region represents the original, ambiguity-free, $2\pi$ dynamic
range of a single interferometer operated at $T=T_{1}$. Full-quadrature
phase-detection allows for a unique solution for all phases, compared
to ambiguities generated when detecting only the cosine component
(e). (f) Dual-$T$ measurements at constant inertial phase $\phi_{a}=0$,
demonstrating that the noise in both interferometers is highly correlated
with slope $\sim\tau$. In red, the covariance ellipse at 95\% confidence
level.\label{fig:1-dual-T-scheme}}
\end{figure*}

In this work, we achieve a dramatic enhancement of dynamic range on
a single-shot basis by combining two powerful approaches in atom interferometry:
increasing the dynamic range without sensitivity loss through small
variations of the interferometer scale factor \citep{Avinadav2019},
and acquiring multiple phase measurements in a single experimental
run \citep{Bonnin2018,Yankelev2019}. First, when the same fundamental
physical quantity determines two interferometric phases with slightly
different scale factors, it can be uniquely extracted within an enhanced
dynamic range, determined by a moiré wavelength which is inversely
proportional to the difference between scale-factors {[}Fig.\,\ref{fig:1-dual-T-scheme}(a){]}.
Second, by operating and reading out the two interferometers simultaneously
within the same experimental shot, major common-mode noises are rejected,
increasing the scheme's robustness to dominant sources of noise. Additionally,
such operation maintains the original temporal bandwidth of the measurement.
Further exponential increase in dynamic range, at the cost of a linear
reduction of temporal bandwidth, is achieved by varying the scale-factor
ratio between shots.

\section*{Principles of Dual-$T$ Interferometry}

We realize the above concept in a Mach-Zehnder atom interferometer
measuring the local acceleration of gravity \citep{Kasevich_1991}.
Such devices use light-pulses as ``atom-optics'' that split the
atomic wavepacket into two ams and later recombine them after they
traveled on macroscopically distinct trajectories. The differential
phase accumulated between the arms of the interferometer depends on
the motion of the atoms.

In our experiment, laser-cooled $^{87}\textrm{Rb}$ atoms are launched
vertically on a free-fall trajectory. Counter-propagating, vertical
laser beams at $\unit[780]{nm}$ drive two-photon Raman transitions
between two electronic ground states while imparting recoil of two
photon momenta \citep{Kasevich1991a}. The Raman beams are sent from
the top and are retro-reflected from a stabilized mirror at the bottom,
which defines the reference frame with respect to which the motion
of the atoms is measured. The interferometric sequence is composed
of three Raman pulses, equally spaced by time $T$, acting to split
the atomic wavepacket into two components that drift apart, and then
to redirect and recombine them, leading to a final atomic population
ratio determined by the phase difference between the two arms.

In this configuration, the phase difference is determined by the gravitational
acceleration $g$ according to $\phi_{a}=\left(k_{\textrm{eff}}g-\alpha\right)T^{2}$,
with $\hbar k_{\textrm{eff}}$ the total momentum transferred by the
Raman interaction, and $\alpha$ a chirp rate applied to the relative
frequency between the Raman beams to compensate for the changing Doppler
shift of the falling atoms. Residual vibrations of the mirror contribute
noise to the inertial phase $\phi_{a}$.

The concept we develop relies on a so-called dual-$T$ operation of
the interferometer. Instead of one pulse sequence, two interleaved
pulse sequences with slightly different $T$ values are performed
{[}Fig.\,\ref{fig:1-dual-T-scheme}(b){]}. By tuning their two-photon
Doppler-detunings, each set of pulses addresses a different vertical
velocity class of the atoms. We operate the two interferometers with
scale factors differing by the ratio $\tau\lesssim1$, choosing the
interferometric durations $T_{1}=T$ and $T_{2}=\sqrt{\tau}T$, with
$T=\unit[55]{ms}$ (see Methods).

\begin{figure*}[t]
\begin{centering}
\includegraphics[width=2\columnwidth]{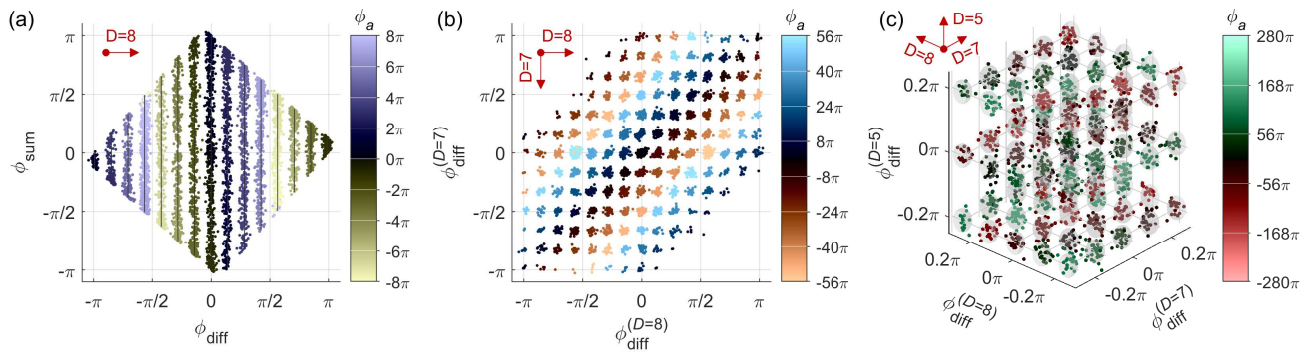}
\par\end{centering}
\centering{}\caption{Analysis of dual-$T$ interferometry measurements in single-shot,
two-shot, and three-shot operation. (a) Single-shot dual-$T$ operation
with $D=8$ for $\phi_{a}$ in the range of $\pm8\pi$ (color coded).
Every dot corresponds to a single measurement. Each discrete value
of $\phi_{\textrm{diff}}$ corresponds to a different sub-range of
$\phi_{a}$, and within that sub-range $\phi_{\textrm{sum}}$ changes
continuously and linearly with $\phi_{a}$. (b) Sequential two-shot
dual-$T$ operation with $D=7,8$ for $\phi_{a}$ in the range of
$\pm56\pi$, presented in the $\phi_{\textrm{diff}}^{\left(D=7\right)}$-$\phi_{\textrm{diff}}^{\left(D=8\right)}$
plane. The discrete clusters in this plane correspond to different
sub-ranges of $\phi_{a}$. (c) Sequential three-shot dual-$T$ operation
with $D=5,7,8$ for $\phi_{a}$ in the range of $\pm280\pi$, presented
in the $\phi_{\textrm{diff}}^{\left(D=5\right)}$-$\phi_{\textrm{diff}}^{\left(D=7\right)}$-$\phi_{\textrm{diff}}^{\left(D=8\right)}$
space. For clarity, only a subset of the solutions around $\phi_{\textrm{diff}}^{(D=5,7,8)}=0$
is presented, and gray ovals surround the expected solutions.\label{fig:2-phase-maps}}
\end{figure*}

Conventionally, the population ratio between the interferometer states
is measured directly, and the cosine of the phase is extracted. In
our dual-$T$ scheme, we detect the phases $\phi_{1},\phi_{2}$ of
both interferometers by acquiring an image of the atoms in one of
the final atomic states. The independent readout of both interferometers
is enabled by the ballistic expansion of the cloud, which maps the
different velocity classes onto different vertical positions. To obtain
the phase, in a manner equivalent to full quadrature detection where
both sine and cosine components of the phase are measured, we use
phase-shear readout \citep{Sugarbaker2013}. We tilt the retro-reflecting
Raman mirror by a small angle before the final $\pi/2$-pulses to
generate a spatial transverse interference pattern across the cloud,
as utilized in point-source interferometry \citep{Dickerson2013,Chen2019,Avinadav2020}
and shown in Fig.\,\ref{fig:1-dual-T-scheme}(c). The phase offset
of this pattern can be directly extracted with constant sensitivity
for all interferometric phases.

Figure \ref{fig:1-dual-T-scheme}(d) shows single-shot measurements
in a dual-$T$ operation with the dynamic range enhanced by a factor
of 8. We vary $\phi_{a}$ by changing the chirp rate $\alpha$ with
respect to its nominal value $\alpha_{0}=k_{\textrm{eff}}g$, thereby
emulating changes in $g$. We find that $\phi_{a}$ is mapped onto
a unique set of straight, parallel lines in the plane spanned by $\phi_{1}$
and $\phi_{2}$ owing to the quadrature detection capability. Conversely,
conventional detection which resolves only the cosine of the phase,
would result in many phase ambiguities due to very different values
of $\phi_{a}$ being mapped to similar measured phase components {[}Fig.\,\ref{fig:1-dual-T-scheme}(e){]},
severely limiting the benefits of a dual-$T$ operation. Quadrature
detection, together with the strong suppression of common noise due
to operation at very similar scale factors, allows the dual-$T$ scheme
to achieve a significantly larger enhancement compared to past implementations
of simultaneous atom interferometers with different scale factors
\citep{Bonnin2018}.

\section*{Dual-$T$ Phase Analysis}

\textbf{Phase estimation for single shot dual-$T$.} The measured
interferometric phases $\phi_{1},\phi_{2}$ are constrained to the
bare dynamic range $\pm\pi$ and can be written as
\begin{align}
\phi_{1} & =\phi_{a}-2\pi n_{1},\label{eq:phi1_simple}\\
\phi_{2} & =\tau\phi_{a}-2\pi n_{2}.\label{eq:phi2_simple}
\end{align}
The integers $n_{1}$ and $n_{2}$, which respectively bring $\phi_{1}$
and $\phi_{2}$ to the range $\pm\pi$, are \emph{a priori} unknown.

We define $D\equiv\left(1-\tau\right)^{-1}$, with $\tau=\left(T_{2}/T_{1}\right)^{2}$
the scale-factors ratio. For integer values of $D$, the dynamic-range
enhancement is exactly $D$; as illustrated in Fig.\,\ref{fig:1-dual-T-scheme}(a),
$\phi_{1}$ and $\phi_{2}$ have a joint period of $2D\pi$ as in
a moiré effect, resulting in an extended ambiguity-free dynamic range
of $\pm D\pi$ (see Methods for discussion on non-integer values).

\begin{figure*}
\begin{centering}
\includegraphics[width=2\columnwidth]{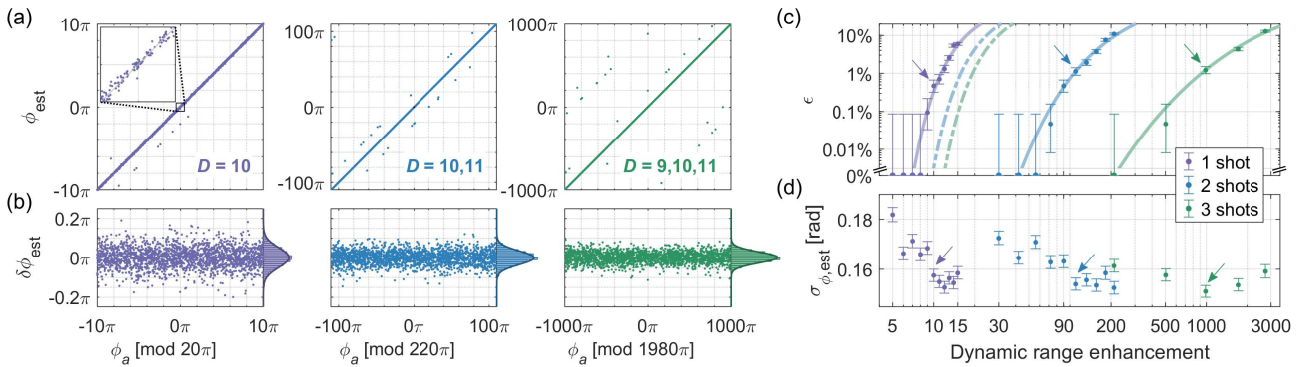}
\par\end{centering}
\centering{}\caption{Performance analysis of dual-$T$ interferometry. (a) Estimated inertial
phase for single-shot measurements (left; inset shown zoom on $\pm\pi/2$
region) and for sequential two-shot (center) and three-shot (right)
measurements, with dynamic range enhancement factors of $10$, $110$,
and $990$, respectively. Outlying measurements appear as data points
visibly distant ($>2\pi$) from their expected value. We observe only
10, 24, and 26 such outliers out of 2000 data points in the three
data sets, respectively. (b) Residuals of the estimated phases, including
only non-outlying measurements. Compared to single-shot measurements,
standard deviations of two- and three-shot residuals are smaller by
factors of $\sqrt{2}$ and $\sqrt{3}$, respectively. (c) Outlier
probability $\epsilon$ as a function of dynamic range enhancement
obtained for individual values of $D$ and for various combinations
of consecutive coprime $D$ values. For $D\protect\leq8$, there were
no outliers in the measured data set. Error bars represent 67\% confidence
intervals of the estimated value. Solid lines are calculated using
Eq.\,(\ref{eq:error_probability}) with $\sigma_{\textrm{ind}}=\unit[80]{mrad}$.
Dashed lines represent outlier probability for an alternative scheme
of averaging two or three sequential shots using a single $D$ value.
(d) Estimation error of gravitation phase per shot $\sigma_{\phi,\textrm{est}}$.
The error is dominated by vibration-induced phase noise and is nearly
equal for all realizations. In (c),(d), arrows indicate the measurements
shown in (a),(b).\label{fig:benchmark}}
\end{figure*}

To analyze a dual-$T$ measurement, we define the quantities $\phi_{\textrm{diff}}$
and $\phi_{\textrm{sum}}$, 
\begin{align}
\left(\begin{array}{c}
\phi_{\textrm{diff}}\\
\phi_{\textrm{sum}}
\end{array}\right) & =\frac{1}{1+\tau^{2}}\left(\begin{array}{cc}
\tau & -1\\
1 & \tau
\end{array}\right)\left(\begin{array}{c}
\phi_{1}\\
\phi_{2}
\end{array}\right).\label{eq:rot_mat}
\end{align}
$\phi_{\textrm{diff}}$ and $\phi_{\textrm{sum}}$ act as coarse and
fine measurements, respectively. As shown in Fig.\,\ref{fig:2-phase-maps}(a),
which presents an analysis of $D=8$ dual-$T$ measurements, $\phi_{\textrm{diff}}$
takes on a discrete set of $2D-1$ values. This constrain uniquely
determines the values of $n_{1}$ and $n_{2}$ and hence the $2\pi$
sub-range in which $\phi_{a}$ lies. Correspondingly, $\phi_{\textrm{sum}}$
is a continuous variable, providing the estimation of the inertial
phase $\phi_{a}$ within that sub-range (see Methods).

\textbf{Phase estimation for sequential operation.} We now turn to
discuss further enhancement of dynamic range obtained by a sequence
of several dual-$T$ shots with alternating integer values of $D$.
Here we fix $T_{1}$ and alternate $T_{2}$ between shots. Assuming
that changes in $\phi_{a}$ are small between consecutive shots, the
above analysis per shot provides $n_{1}\,\text{mod}\,D$. Taken together,
the full sequence uniquely determines $n_{1}$ within a range defined
by the least common multiple of the employed $D$ values, or, for
coprime integers, simply their product (see Methods and Fig.\,\ref{fig:supp-ruler}).

Analyses of two-shot operation with $D=7,8$ and three-shot operation
with $D=5,7,8$ are shown in Fig.\,\ref{fig:2-phase-maps}(b,c).
Each data point is a measurement with a random value of $\phi_{a}$
within the extended dynamic ranges $\pm56\pi$ and $\pm280\pi$, respectively.
We observe two- and three-dimensional clustering of the differential
phases $\phi_{\textrm{diff}}$, where each cluster corresponds to
a unique, non-ambiguous phase range smaller than $2\pi$.

\textbf{Noise and outlier probability.} By virtue of simultaneously
operating the two interferometers with similar scale factors, vibrations-induced
phase noise is highly correlated between them {[}Fig.\,\ref{fig:1-dual-T-scheme}(f){]}
and has negligible contribution to $\phi_{\textrm{diff}}$. The dominant
noise in $\phi_{\textrm{diff}}$ results from uncorrelated, independent
detection noise in $\phi_{1}$ and $\phi_{2}$, whose standard deviation
we denote as $\sigma_{\textrm{ind}}$ (see Methods for a detailed
discussion of noise terms).

As $D$ is increased, and the discrete values of $\phi_{\textrm{diff}}$
become denser, the uncorrelated noise may lead to errors in determining
the correct sub-range for $\phi_{a}$, producing an outlier with phase
estimation error in multiples of $2\pi$. The probability $\epsilon$
for a measurement to be such an outlier is approximately {[}see Eq.\,(\ref{eq:precise-err-prob})
for exact expression{]}
\begin{equation}
\epsilon\approx\mathrm{erfc}\left(\frac{\pi}{2}\frac{1}{D\cdot\sigma_{\textrm{ind}}}\right).\label{eq:error_probability}
\end{equation}
Crucially, $\epsilon$ depends only on the uncorrelated noise and
not on the vibrations-dominated correlated noise, which is typically
much larger. In the data presented in Fig.\,\ref{fig:2-phase-maps}(a),
we observe one such outlier out of 5000 measurements for $D=8$.

For the case of sequential dual-$T$ operation, the total outlier
probability $\epsilon_{\textrm{seq}}$ depends on the outlier probabilities
in each shot and, in the relevant regime of small error probabilities,
is given simply by their sum. For any desired dynamic range and temporal
bandwidth, the outlier probability is minimized by choosing \emph{consecutive}
coprime values of $D$.

\section*{Experimental Characterization}

\textbf{Performance analysis.} To quantify the performance of the
dual-$T$ scheme in terms of phase sensitivity and outlier probability,
we extend the phase scan to random, known, values of $\phi_{a}$ within
the range of $\pm1000\pi$, corresponding to accelerations of $\unit[\pm65]{mm/s^{2}}$
at $T=\unit[55]{ms}$. For each phase, we perform measurements with
$D$ values between 5 and 15, and perform dual-$T$ analysis using
each $D$ separately, using pairs of consecutive $D$ values, and
using triplets of consecutive coprime $D$ values. We analyze each
measurement within its appropriate extended dynamic range; data points
that are outside the measurement's relevant dynamic range are wrapped
back onto it. We then compare the extracted phase to its expected
value, from which we estimate the outlier probability $\epsilon$
as well as the phase residuals of the measurements without outliers.

The results, presented in Fig.\,\ref{fig:benchmark}, demonstrate
an enhancement of dynamic range by factors of $10$ in a single shot,
$\sim100$ in two shots, and $\sim1000$ in three shots, while maintaining
phase residuals of $\sigma_{\textrm{\ensuremath{\phi,}est}}\sim\unit[160]{mrad/shot}$
($\unit[\sim3.3]{\mu m/s^{2}/shot}$), and with outlier probabilities
of 0.5\%, 1.1\%, and 1.2\%, respectively. In general, we find excellent
agreement with the error model described by Eq.\,(\ref{eq:error_probability}),
with $\sigma_{\textrm{ind}}=\unit[80]{mrad}$ estimated from these
data.

We note that an outlier fraction on the order of 1\% is acceptable
in most applications, as such outliers can be identified and removed
by comparison to adjacent shots or using auxiliary measurements. However,
even if nearly zero outlier fraction is required, the dual-$T$ scheme
can deliver a significant dynamic range enhancement. For example,
with the above measured value of $\sigma_{\textrm{ind}}$ and for
$D=6$, we expect $\epsilon\approx3\times10^{-6}$.

Furthermore, averaging over $N$ repeated measurements with the same
$D$ value can decrease the outlier probability $\epsilon$ by effectively
reducing $\sigma_{\textrm{ind}}$ by a factor $\sqrt{N}$. However,
by employing the same number of sequential measurements with \emph{alternating}
values of $D$ as described above, the same value of $\epsilon$ may
be achieved with significantly larger dynamic range enhancement, as
seen from comparing solid and dashed curves in Fig.\,\ref{fig:benchmark}(c).

\begin{figure}[!t]
\begin{centering}
\includegraphics[bb=0bp 0bp 396bp 285bp,width=1\columnwidth]{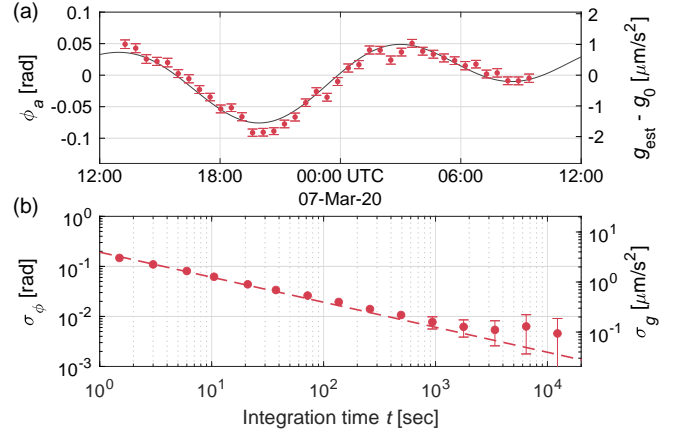}
\par\end{centering}
\centering{}\caption{Stability of dual-$T$ interferometry. (a) Time series of $\phi_{\textrm{sum}}$,
with half-hour binning, measured with $D=10$. The results follow
the expected tidal gravity variation as calculated from solid-earth
model (black solid line). (b) Allan deviation of the residuals of
$\phi_{\textrm{sum}}$ from the tidal model. Dashed line is a fit
to $t^{-1/2}$ with sensitivity per shot of $\unit[155]{mrad}$ ($\unit[3.2]{\mu m/s^{2}}$).\label{fig:stability}}
\end{figure}

\textbf{Stability of dual-$T$ interferometry.} To demonstrate the
long-term stability of dual-$T$ interferometry, we continuously measure
gravity over 20 hours with $D=10$. As shown in Fig.\,\ref{fig:stability},
$\phi_{a}$ follows the expected tidal gravity variations throughout
the measurement period. It remains stable at time scales of $\unit[10^{4}]{sec}$,
to better than $\unit[100]{nm/s^{2}}$, showing that the dual-$T$
scheme does not add significant drifts to the estimated phase. Conversely,
$\phi_{\textrm{diff}}$ does exhibit small drifts which we attribute
to mutual light-shift between the two interferometers. However, due
to the discrete nature of $\phi_{\textrm{diff}}$, these drifts can
be easily corrected in several ways (see Methods).

\section*{Tracking fast-varying signals}

We now turn to discuss dynamic scenarios, such as mobile gravity surveys
or inertial measurements on a navigating platform, where the measured
acceleration and thus $\phi_{a}$ change dramatically between shots.
Dual-$T$ interferometry with fixed $D$ can directly track a signal
that randomly varies by up to $\pm D\pi$ from shot to shot. Moreover,
alternating the value of $D$ between consecutive measurements can
enable tracking a signal with even larger variations; however, the
sequential analysis described above cannot be applied due to the phase
changing between shots, and a different analysis method is required.

\begin{figure}[!t]
\begin{centering}
\includegraphics[width=1\columnwidth]{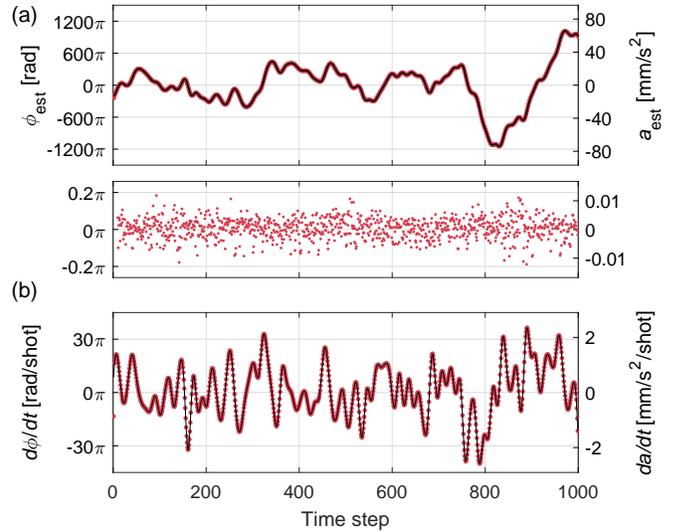}
\par\end{centering}
\centering{}\caption{Tracking of a time-varying acceleration using dual-$T$ interferometry
combined with a particle-filter protocol. Measurements are performed
with alternating $D=9,10$. (a) Acceleration signal extracted from
the measurements using the particle filter (red), compared to the
input signal (black). Bottom panel shows the residuals with standard
deviation of $\unit[174]{mrad}$ ($\unit[3.6]{\mu m/s^{2}}$). (b)
Temporal derivative (shot-to-shot variation) of the measured acceleration
signal (red) compared to the input (black).\label{fig:particle}}
\end{figure}

To track such a signal, we employ a particle filter estimation protocol
\citep{DelMoran1996,van2001unscented}. Particle filtering is a powerful
and well-established technique in navigation science, signal processing
and machine learning, among other fields. It is a sequential, Monte-Carlo
estimation approach based on a large number of particles which represent
possible hypotheses of the system's current state, \emph{e.g.}, the
inertial phase measured by the sensor. These hypotheses are weighted
through Bayesian estimation after every measurement, converging on
a solution that is consistent with the sensor readings over time.
In our context, under some model assumptions on the signal dynamics,
use of particle filter enables full recovery of the single-shot bandwidth
\citep{Avinadav2020} while maintaining the large increase in dynamic
range rendered by the sequential operation.

An experimental realization of tracking a dynamic signal is presented
in Fig.\,\ref{fig:particle}. We change the chirp rate $\alpha$
between shots to simulate a band-limited random walk of $a$ and perform
dual-$T$ measurements with alternating $D=9,10$. The sequence of
measured phases is then analyzed with a particle filter protocol using
a second-order derivative model (see Methods), to extract best estimate
for the time-series of $\phi_{a}$. Following a brief convergence
period (Supplementary Fig.\,\ref{fig:supp-phi-diff}), we successfully
track this time-varying signal which spans over $2000\pi$ and changes
by up to $40\pi$ between shots, with sensitivity per shot similar
to measurements of static signals under similar conditions, and with
no outliers. We note that while the analysis was carried out in post-process,
it is in principle compatible with implementation as a real-time protocol.

\section*{Discussion}

In conclusion, we present a novel approach to atom interferometry
for significant enhancement of dynamic range without a reduction in
sensitivity and with high measurement bandwidth. In applications where
traditional atom interferometers must be operated at reduced sensitivity
due to the expected dynamic range of the measured signal, our approach
enables measurements with a substantial increase in sensitivity while
maintaining the necessary dynamic range.

Taking advantage of full-quadrature phase detection and common-noise
rejection, we experimentally demonstrate an increase of dynamic range
by more than an order of magnitude in a single shot. Incorporating
data from several consecutive shots, the dynamic range further increases
in exponential fashion, allowing us to reach three orders of magnitude
gain using only three measurements. Finally, we demonstrate tracking
of a dynamical signal with tens of radians shot-to-shot variation
by combining the dual-$T$ measurement with a particle-filter protocol,
representing a major improvement compared to recent works \citep{Avinadav2019}.

This approach can dramatically enhance performance of sensors, and
in particular inertial-sensing atom interferometers, under challenging
conditions, by enabling non-ambiguous operation without sacrificing
either sensitivity or bandwidth. Such conditions are encountered in
field operation of such sensors, for example in mobile gravity surveys
or when used for inertial navigation on a moving platform. By extending
the sensor real-time dynamic range, the requirements on vibration
isolation or corrections based on auxiliary measurements can be relaxed,
or equivalently, existing sensors can be operated in more demanding
environments.

Dual-$T$ measurements can be realized by multiple means, based on
known atom interferometry tools, \emph{e.g.}, dual-species interferometry
\citep{Bonnin2018} or momentum-state multiplexing \citep{Yankelev2019},
in addition to phase-shear readout \citep{Sugarbaker2013} used in
this work. It is also compatible with important atom-interferometry
practices, such as $k$-reversal \citep{McGuirk2002,Louchet-Chauvet2011}
and zero-dead-time operation \citep{Dutta2016}. Further improvement
of the scheme is possible by incorporating more than two interferometric
sequences within the same experimental shot, enabling the gain demonstrated
here for sequential operation within a single shot. Extension of the
approach to other atom interferometry configurations is also possible.

\section*{Acknowledgments}

This work was supported by the Pazy Foundation, the Israel Science
Foundation, and the Consortium for quantum sensing of Israel Innovation
Authority.

\bibliography{../_main_library}

\smallskip{}

\onecolumngrid
\begin{center}
\par\end{center}

\newpage\smallskip{}

\appendix
\twocolumngrid

\section*{Methods}


\textbf{Experimental sequence.} We load a cloud of $^{87}\textrm{Rb}$
atoms in a magneto-optical trap (MOT) and launch it upwards at $\unit[0.9]{m/s}$
with moving optical molasses, which also cools the cloud to $\unit[5]{\mu K}$.
Atoms initially populate equally all $m_{F}$ sub-levels in the $F=2$
hyperfine manifold. We select atoms in two distinct velocity classes
and in the $m_{F}=0$ state using two counter-propagating Raman $\pi$-pulses,
with $\unit[20]{\mu s}$ duration and a relative Doppler detuning
of $\unit[80]{kHz}$. Two interferometric sequences of $\pi/2$-$\pi$-$\pi/2$
pulses, with durations of $12$, $24$, and $\unit[12]{\mu s}$ respectively,
address each of the velocity classes as shown in Fig.\,\ref{fig:1-dual-T-scheme}.
The timing of the $\pi$ pulses of the two interferometers is set
to $\unit[22]{ms}$ and $\unit[22.5]{ms}$ after the apex of the trajectories.
The precise ratio of $T_{2}/T_{1}$ contains empirically-calibrated
corrections on the order of $10^{-5}$ with respect to the naive $\sqrt{\tau}$
value, attributed mainly to finite Raman pulse durations \citep{Fang2018}.
Before the final $\pi/2$-pulses, the Raman mirror is tilted by $\unit[120]{\mu rad}$.
With the MOT beams tuned on resonance with the $F=2\rightarrow F=3$
cycling transition, a fluorescence image of atoms in the $F=2$ level
is taken on a CCD camera oriented perpendicularly to the Raman mirror
tilt axis. The experiment is repeated every 2 to 3 seconds.

\textbf{Extraction of the measured phases $\phi_1,\phi_2$.} We first
integrate the image horizontally to find the vertical Gaussian envelopes
of the fringe patterns, which are used to define the analysis region-of-interest
for each interferometer (Supplementary Fig.\,\ref{fig:supp-image}).
We then vertically integrate the image over those regions and fit
the resulting profile to Gaussian envelopes with sinusoidal modulation.
The phases of the measurement are taken as the phases of the fitted
fringes at the horizontal center of the cloud. Finally, we calculate
and correct the vibration-induced phase based on the auxiliary accelerometer
signal, taking into account the different interrogation times of each
interferometer.

\textbf{Single-shot dual-$T$ analysis.} For each dual-$T$ shot,
we rotate the measured $\phi_{1},\phi_{2}$ according to Eq.\,(\ref{eq:rot_mat}),
\begin{align}
\phi_{\textrm{diff}} & =\frac{2\pi}{1+\tau^{2}}\left(\frac{n_{1}}{D}-\Delta n\right),\label{eq:phi_diff_def}\\
\phi_{\textrm{sum}} & =\phi_{a}-2\pi\frac{\left(n_{1}+\tau n_{2}\right)}{1+\tau^{2}}.
\end{align}
Within the extended dynamic range of $\pm D\pi$ for $\phi_{a}$,
the integer $n_{1}$ takes values within $\pm\left\lfloor D/2\right\rfloor $,
and $\Delta n=n_{1}-n_{2}$ takes either $0,\pm1$. From $\phi_{\textrm{diff}}$
we uniquely determine $\Delta n$,

\begin{align}
\Delta n & =\begin{cases}
0 & \left|\phi_{\textrm{diff}}\right|<\frac{\pi}{1+\tau^{2}}\left(1-\frac{1}{2D}\right)\\
-\mathrm{sgn}\left(\phi_{\textrm{diff}}\right) & \left|\phi_{\textrm{diff}}\right|>\frac{\pi}{1+\tau^{2}}\left(1-\frac{1}{2D}\right)
\end{cases},
\end{align}
and $n_{1}$ follows as the round value of $D\left[\left(1+\tau^{2}\right)\phi_{\textrm{diff}}/\left(2\pi\right)+\Delta n\right]$.
Finally, we estimate $\phi_{a}$ by substituting $n_{1}$ and $n_{2}=n_{1}-\Delta n$
back into $\phi_{\textrm{sum}}$.

We focused the discussion on integer $D$. Rational $D$ yields joint
phase periodicity according to the lowest term numerator of $D$,
but with less efficient common-mode noise rejection. For irrational
$D$, there is no well-defined periodicity and hence no discrete set
of allowed $\phi_{\textrm{diff}}$ solutions. While in both cases
dynamic range enhancement is attained, optimal results are achieved
for integer $D$.

\textbf{Sequential dual-$T$ analysis.} From a sequence of $N$ ($N=2,3$
in this work) shots with alternating $D^{(i)}$, where $i=1,\ldots,N$,
we retrieve $N$ pairs of phases $[\phi_{1}^{\left(i\right)},\phi_{2}^{\left(i\right)}]$.
Analyzing each shot separately as described above, we extract from
them a set of values $\tilde{n}_{1}^{(i)}$, each within $\pm\left\lfloor D^{\left(i\right)}/2\right\rfloor $.
Joint analysis of the sequential measurements in principle amounts
to finding the integer $n_{1}$ that satisfies the set of equations
$\tilde{n}_{1}^{(i)}=n_{1}\,\text{mod}\,D^{\left(i\right)}$. The
solution is unique within the range $\pm\text{LCM}\left(D^{\left(1\right)},\ldots,D^{\left(N\right)}\right)$,
$\text{LCM}$ denoting the least common multiple. This analysis assumes
that $n_{1}^{(i)}=n_{1}^{(1)}$ for all $i$, as the first interferometer
always measures $\phi_{a}$ with the same interrogation time $T$.
However, for values of $\phi_{a}$ close to odd multiples of $\pi$,
phase noise may cause variations of up to $\pm1$ in $n_{1}^{\left(i\right)}$.
We calculate the variations $\Delta n_{1}^{\left(i\right)}=n_{1}^{(1)}-n_{1}^{(i)}$
for $i>1$ as the round value of $(\phi_{1}^{(1)}-\phi_{1}^{(i)})/\left(2\pi\right)$
and take them into account when solving the set of equations described
above for $n_{1}$. In Fig.\,\ref{fig:2-phase-maps}(b), only measurements
with $\Delta n_{1}^{\left(2\right)}=0$ are shown for clarity; the
full range of results is shown in Supplementary Fig.\,\ref{fig:supp-sequential}.

\textbf{Experimental noise parametrization.} Extending on Eqs. (\ref{eq:phi1_simple}-\ref{eq:phi2_simple}),
we write the phases $\phi_{1},\phi_{2}$ as
\begin{align}
\phi_{1} & =\left(\phi_{a}+\delta\phi_{\textrm{corr}}\right)+\delta\phi_{\textrm{ind},1}-2\pi n_{1},\\
\phi_{2} & =\tau\left(\phi_{a}+\delta\phi_{\textrm{corr}}\right)+\delta\phi_{\textrm{ind},2}-2\pi n_{2}.
\end{align}
Here $\delta\phi_{\textrm{corr}}\sim\mathcal{N}\left(0,\sigma_{\textrm{corr}}^{2}\right)$
is the noise term on the inertial signal common to both interferometers,
whereas $\delta\phi_{\textrm{ind},1},\delta\phi_{\textrm{ind},2}\sim\mathcal{N}\left(0,\sigma_{\textrm{ind}}^{2}\right)$
are independent noise terms, \emph{e.g.}, due to detection noise of
each interferometer. While the methodology and data processing will
work well for any noise covariance, this parametrization is natural
to operating the two interferometers simultaneously and with similar
scale factors, such that the inertial noise is highly correlated as
demonstrated in Fig.\,\ref{fig:1-dual-T-scheme}(f).

With this parametrization and based on Eq.\,(\ref{eq:rot_mat}),
$\phi_{\textrm{diff}}$ and $\phi_{\textrm{sum}}$ are characterized
by random noise with standard deviations $\sigma_{\textrm{ind}}/\sqrt{1+\tau^{2}}$
and $\sigma_{\phi,\textrm{est}}=\sqrt{\sigma_{\textrm{corr}}^{2}+\sigma_{\textrm{ind}}^{2}/\left(1+\tau^{2}\right)}$,
respectively. An outlier measurement occurs when the random deviation
of $\phi_{\textrm{diff}}$ from its theoretical value is larger than
half the difference between its discrete solutions, which is $2\pi/\left[D\left(1+\tau^{2}\right)\right]$.
The probability of such an event is given by 
\begin{equation}
\epsilon=\mathrm{erfc}\left(\frac{\pi}{\sqrt{2}}\frac{1}{D\sqrt{1+\tau^{2}}}\frac{1}{\sigma_{\textrm{ind}}}\right),\label{eq:precise-err-prob}
\end{equation}
and approximated by Eq.\,(\ref{eq:error_probability}) for $\tau\approx1$.
See Supplementary Information for experimental noise characterization.

\textbf{Systematic phase shifts.} Dual-$T$ measurements have several
systematic phase shifts which are common also to conventional atom
interferometers, due to factors such as one-photon light shifts, two-photon
light shifts \citep{Gauguet2008}, and offset of the Raman frequency
from Doppler resonance \citep{Gillot2016}. Typically, these effects
are either estimated and accounted for theoretically, or they are
eliminated through wave-vector reversal ($k$-reversal) \citep{McGuirk2002,Louchet-Chauvet2011}.

Nevertheless, some of these shifts may be complicated or modified
by the existence of two simultaneous interferometer pulse sequences,
while new sources of systematic shifts may arise, such as due to an
estimation error of the cloud center position when using phase shear
readout. As demonstrated in Fig.\,\ref{fig:stability}, these effects
do not contribute to bias instability in the measured phase up to
few $\unit{mrad}$, although they may introduce a constant bias which
can be determined and calibrated in advance by comparison of dual-$T$
measurements with standard interferometry. We correct this bias by
performing 15 to 50 initial calibration measurements for different
$D$ values and $k_{\textrm{eff}}$ signs, where we assume prior knowledge
of $\phi_{a}$. For the time-varying experiment in Fig.\,\ref{fig:particle},
these calibration measurements are not included in the particle filter
analysis.

\textbf{Correction of drifts in the differential phase.} As shown
in Fig.\,\ref{fig:supp-phi-diff}(a), $\phi_{\textrm{diff}}$ exhibits
small drifts over time from its expected discrete value. While these
drifts do not directly enter into the estimation of $\phi_{a}$, they
may have a large impact on outlier probability $\epsilon$. By performing
$k$-reversal, we observe that the drift in $\phi_{\textrm{diff}}$
is anti-symmetric with respect to $k_{\textrm{eff}}$. We therefore
attribute the observed phase drifts to differential light-shift between
the two interferometric states of the Raman pulses. As the temporal
response function to external phase-shifts is anti-symmetric with
respect to the central $\pi$-pulse, normally the effect of light
shifts due to the interferometer pulses cancels up to changes in the
light shift during the interferometer due to laser intensity fluctuations
\citep{Cheinet2008}. In our dual-$T$ realization, the light shift
induced by the $\pi/2$-pulses of the shorter interferometer on the
longer one still cancel as before, but each interferometer experiences
an uncompensated light shift due to the $\pi$-pulse of its counterpart.
A realization of dual-$T$ with simultaneous $\pi$-pulses for both
interferometers will circumvent this effect \citep{Bonnin2018}.

These mutual light shifts will be of approximately equal amplitude
but opposite signs, therefore they are suppressed in $\phi_{\textrm{sum}}$
by a factor $\left(1-\tau\right)/\left(1+\tau^{2}\right)$ but amplified
in $\phi_{\textrm{diff}}$ by a factor $\left(1+\tau\right)/\left(1+\tau^{2}\right)$.
These effects of light shifts are entirely canceled by performing
$k$-reversal, and indeed, as we observe in Fig.\,\ref{fig:supp-phi-diff}(b),
the average value of $\phi_{\textrm{diff}}$ over $\pm k_{\textrm{eff}}$
remains stable at time scales of $\unit[10^{4}]{sec}$ to better than
$\unit[1]{mrad}$. In the particle filter demonstration, we used both
$k_{\textrm{eff}}$ signs to correct such drifts, demonstrating the
compatibility of the $k$-reversal technique with the dual-$T$ approach.

Additionally, due to the discrete nature of $\phi_{\textrm{diff}}$,
the observed drifts can also be deterministically corrected without
requiring $k$-reversal, and thus with practically no impact on the
interferometer performance or bandwidth. For the data presented in
Figs.\,\ref{fig:2-phase-maps} and \ref{fig:benchmark}, we continuously
correct drifts in $\phi_{\textrm{diff}}$, without assuming prior
knowledge of $\phi_{a}$, by tracking the difference between the measured
$\phi_{\textrm{diff}}$ values from the nearest discrete values and
subtracting their long-term, moving average.

\textbf{Particle filter implementation.} We choose as state variables
the instantaneous value of the inertial phase $\phi_{a}$ and its
first- and second-order time derivatives, denoting $\boldsymbol{x}^{\left(m,i\right)}=[\begin{array}{ccc}
\phi_{a}^{\left(m,i\right)} & \dot{\phi}_{a}^{\left(m,i\right)} & \ddot{\phi}_{a}^{\left(m,i\right)}\end{array}]^{T}$ for the $m$-th particle at the $i$-th time step. As observables,
we choose the two interferometer phases $\phi_{1,2}$. The initial
value and derivatives of the input $\phi_{a}$ signal are approximately
$-207\pi$, $8.6\pi/dt$, and $4.5\pi/dt^{2}$, respectively. We represent
a scenario where some imperfect knowledge about the starting conditions
exists by drawing the initial values of the particles from normal
distributions characterized by
\begin{equation}
\begin{cases}
\mu_{\phi}^{(0)}=-250\pi & \sigma_{\phi}^{\left(0\right)}=50\pi,\\
\mu_{\dot{\phi}}^{(0)}=0 & \sigma_{\dot{\phi}}^{\left(0\right)}=10\pi/dt,\\
\mu_{\ddot{\phi}}^{(0)}=0 & \sigma_{\ddot{\phi}}^{\left(0\right)}=8\pi/dt^{2}.
\end{cases}
\end{equation}

At each time step of the filter, we first propagate the particles'
state according to $\boldsymbol{x}^{\left(m,i+1\right)}=\mathbf{F}\cdot\boldsymbol{x}^{\left(m,i\right)}+\boldsymbol{w}^{\left(m,i\right)}$,
with $\mathbf{F}$ being the state propagation matrix and $\boldsymbol{w}^{\left(m,i\right)}$
a random process noise with zero mean and covariance matrix $\mathbf{Q}$.
For our model, we have
\begin{equation}
\mathbf{F}=\left[\begin{array}{ccc}
1 & dt & \frac{1}{2}dt^{2}\\
0 & 1 & dt\\
0 & 0 & 1
\end{array}\right],\,\,\mathbf{Q}=dt^{2}\left[\begin{array}{ccc}
0 & 0 & 0\\
0 & 0 & 0\\
0 & 0 & q_{\ddot{a}}^{2}
\end{array}\right],
\end{equation}
where $dt$ is the time interval between consecutive measurements.
Following state propagation, we calculate the expected interferometer
signals for each particle as $\phi_{1}^{\left(m,i\right)}=\phi_{a}^{\left(m,i\right)}$
and $\phi_{2}^{\left(m,i\right)}=\tau^{\left(i\right)}\phi_{a}^{\left(m,i\right)}$,
where $\tau^{\left(i\right)}$ is the scale-factor ratio in the $i$-th
measurement. We calculate their residuals from the actual measurements
modulo $2\pi$ and weigh each particle according to the likelihood
that these residuals are consistent with the independent measurement
noise $\sigma_{\textrm{ind}}$, which we take as $\unit[73]{mrad}$
according to the spread of $\delta\phi_{\textrm{ind,diff}}$. State
variables estimation is achieved by using a ridge-detection algorithm
(MATLAB \texttt{tfridge} function) on the time-dependent particle
histogram to estimate $\phi_{\textrm{est}}^{\left(i\right)}$, as
demonstrated in Supplementary Fig.\,\ref{fig:supp-filter}. For $q_{\ddot{a}}$,
we took a value of $6\pi/dt^{2}$, as it minimizes the mean error
of the estimated $\phi_{a}$ from the measured $\phi_{1,2}$.

\newpage{}

\onecolumngrid
\appendix

\section*{Supplementary Figures}

\setcounter{figure}{0}
\renewcommand{\thefigure}{S\arabic{figure}}

\begin{figure}[H]
\begin{centering}
\includegraphics[bb=0bp 0bp 792bp 210bp,width=1\textwidth]{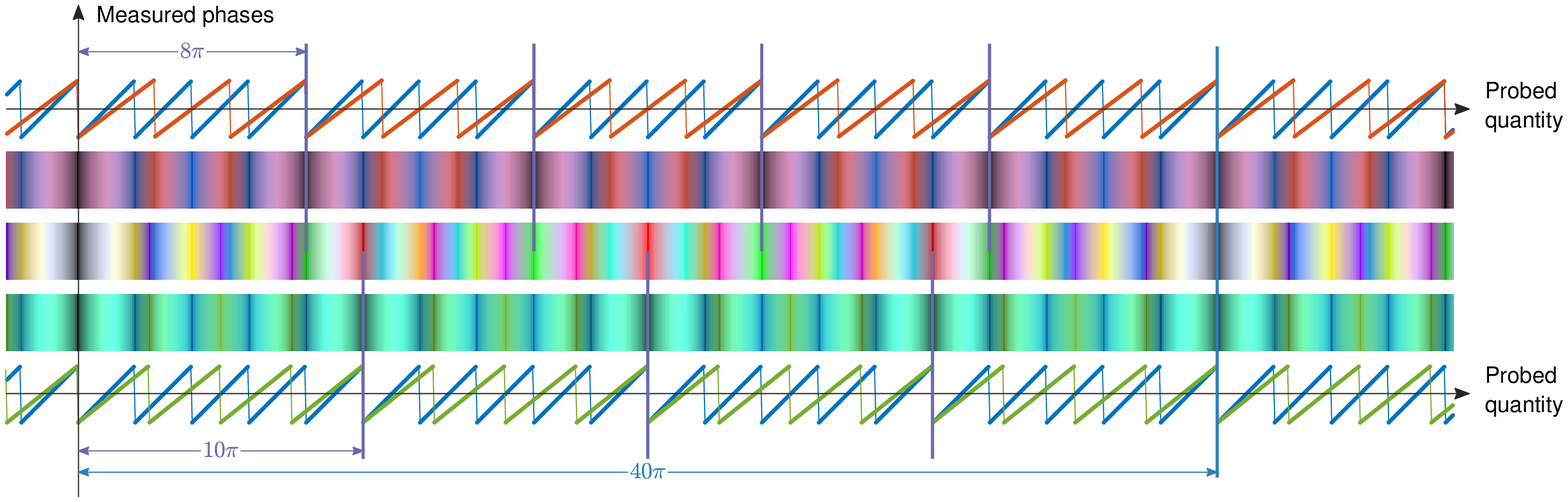}
\par\end{centering}
\centering{}\caption{Principle of dynamic-range enhancement with sequential operation.
In this example, combining dual-$T$ interferometers with enhancement
factors of $\times4$ (top) and $\times5$ (bottom) provides overall
enhancement of $\times20$ in two shots. \label{fig:supp-ruler}}
\end{figure}

\begin{figure}[H]
\begin{centering}
\includegraphics[bb=0bp 0bp 396bp 413bp,width=0.5\textwidth]{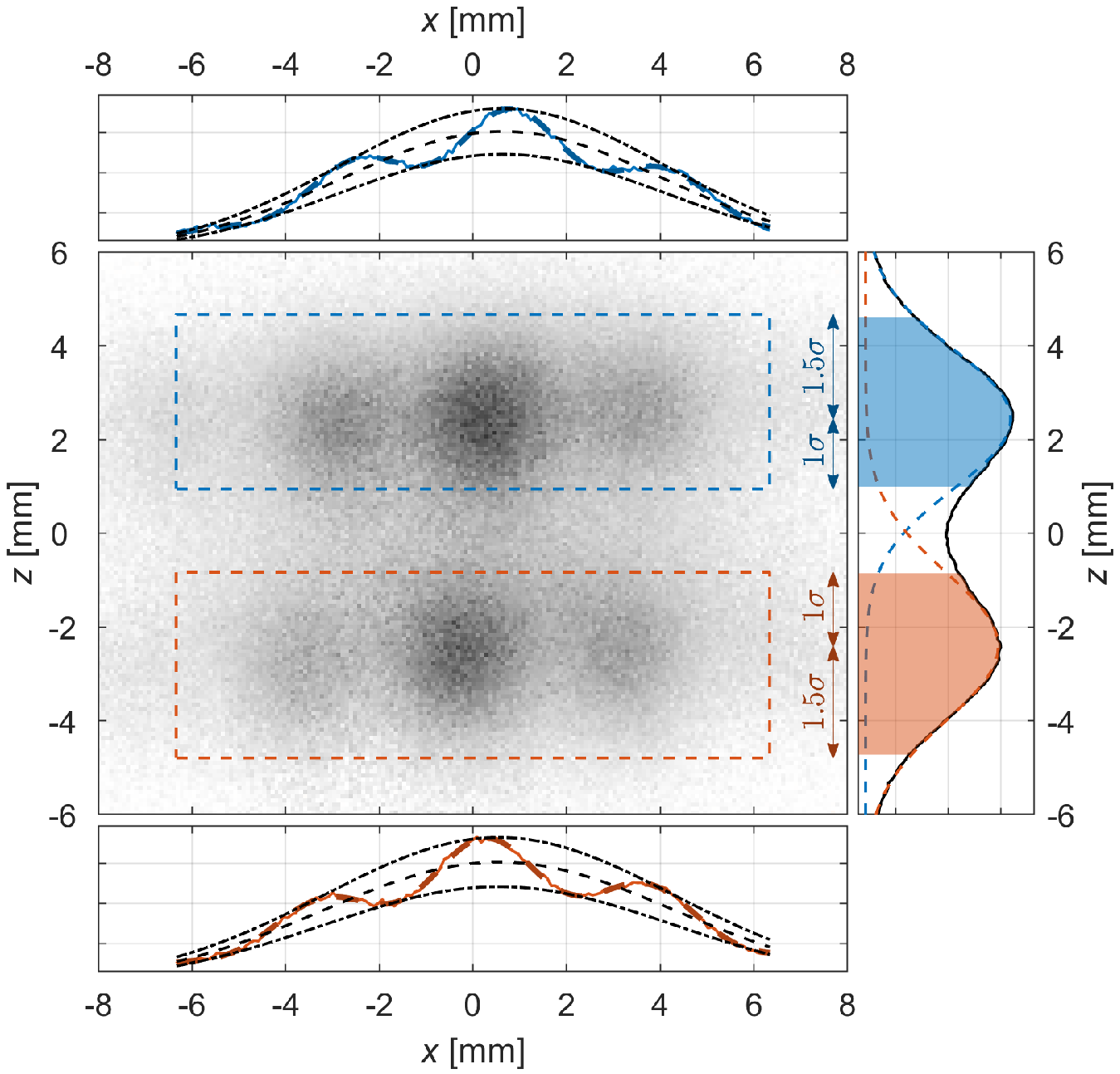}
\par\end{centering}
\centering{}\caption{Dual-$T$ image analysis. Center panel shows raw fluorescence image
and, on the right, the horizontal integration with fit to two Gaussian
envelopes. From the fit, we determine the analysis region-of-interest
for each interferometer (dashed rectangles), over which we integrate
the image vertically and fit to a Gaussian envelope with sinusoidal
modulation (top and bottom panels).\label{fig:supp-image}}
\end{figure}

\begin{figure}[H]
\begin{centering}
\includegraphics[bb=0bp 0bp 792bp 210bp,width=1\textwidth]{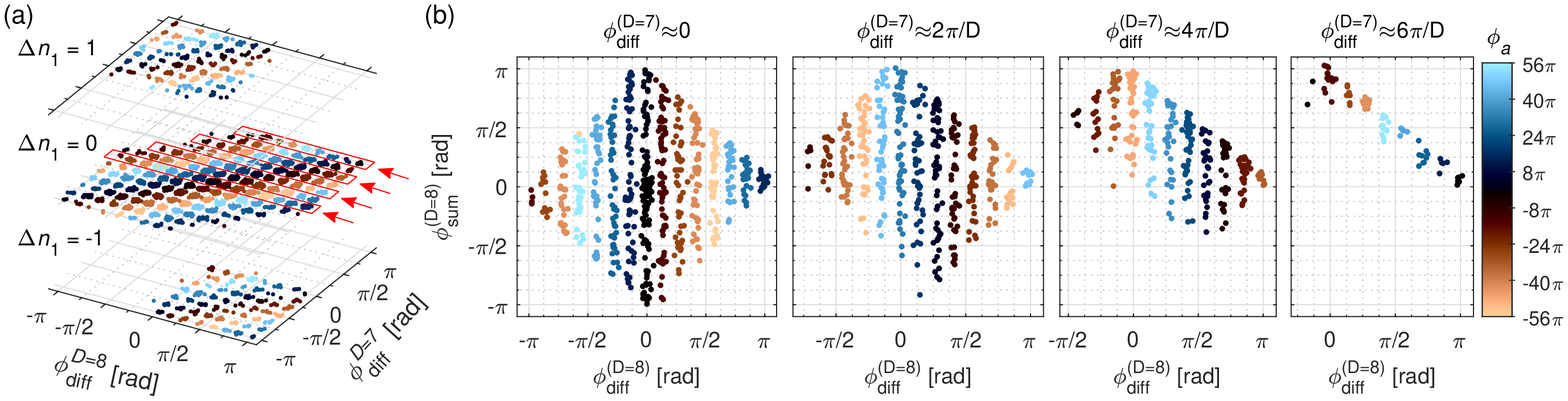}
\par\end{centering}
\centering{}\caption{Extended display of the sequential (two-shot) measurements with $D=7,8$
shown in Fig.\,\ref{fig:2-phase-maps}(b). (a) Results in the $\phi_{\textrm{diff}}^{\left(D=7\right)}$-$\phi_{\textrm{diff}}^{\left(D=8\right)}$
plane, including data points with $\Delta n_{1}\protect\neq0$ which
result from measurements where $\phi_{a}$ is near odd multiples of
$\pi$. Note that $\Delta n_{1}$ is extracted from the measurements
as well, and therefore all these data points are valid sensor measurements.
(b) Results in the $\phi_{\textrm{diff}}^{\left(D=8\right)}$-$\phi_{\textrm{sum}}^{\left(D=8\right)}$
plane, for data points with $\Delta n_{1}=0$ and specific values
of $\phi_{\textrm{diff}}^{\left(D=7\right)}$ as specified in the
title and indicated by red rectangles in (a). \label{fig:supp-sequential}}
\end{figure}

\begin{figure}[H]
\begin{centering}
\includegraphics[bb=0bp 0bp 396bp 413bp,width=0.5\textwidth]{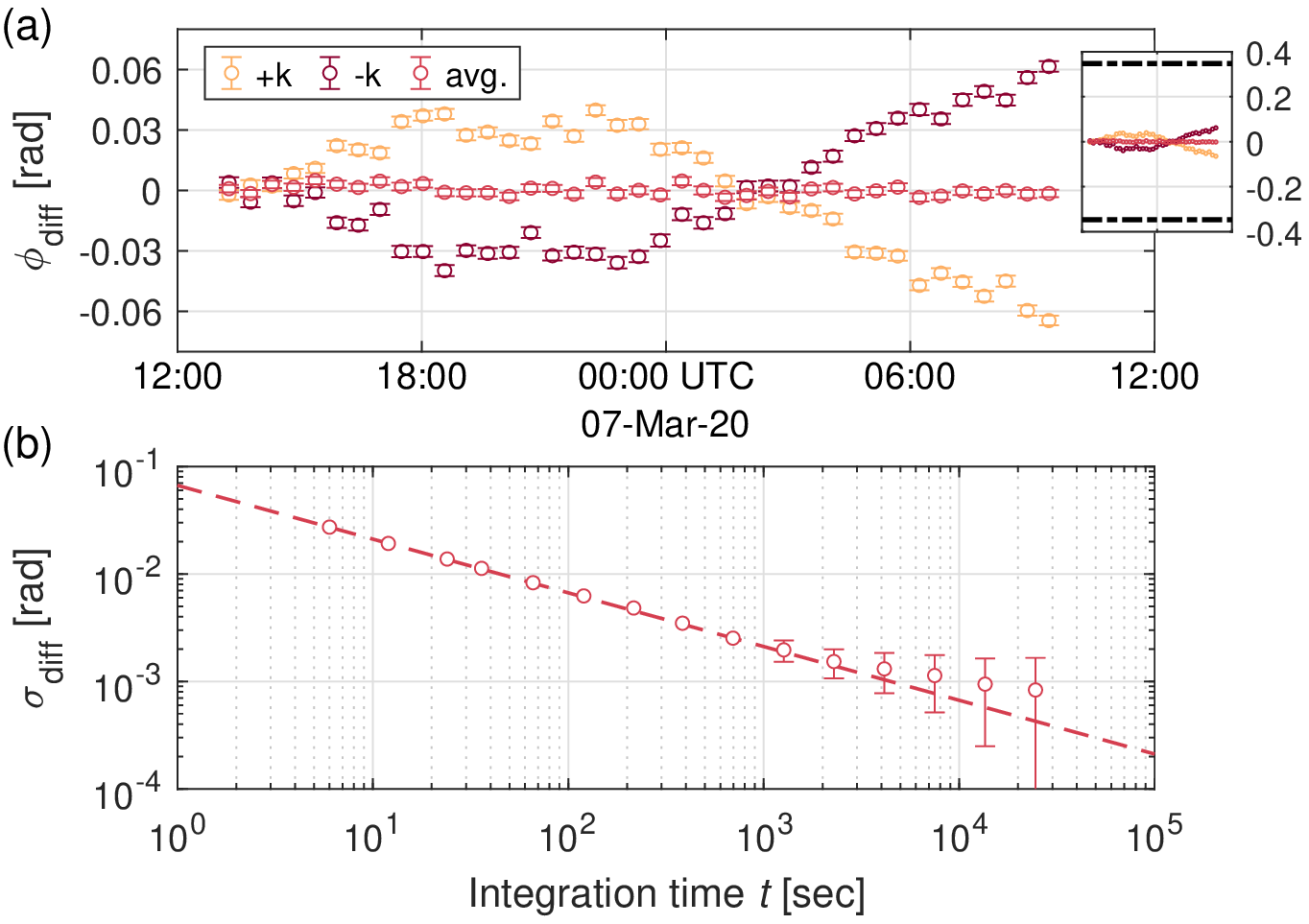}
\par\end{centering}
\centering{}\caption{Analysis of drifts in $\phi_{\textrm{diff}}$ based on the data collected
in the experiment presented in Fig.\,\ref{fig:stability}, with $D=10$.
(a) Time series of $\phi_{\textrm{diff}}$, with half-hour binning.
Systematic phase shifts, which drift by tens of $\unit{mrad}$, are
evident for both signs of $k_{\textrm{eff}}$ but are strongly suppressed
after averaging. The observed drifts are much smaller that the spacing
between the discrete values of $\phi_{\textrm{diff}}$ ($\sim\pm\pi/9\protect\cong\pm\unit[0.35]{rad}$
for $D=10$, see inset). (b) Allan deviation of $\phi_{\textrm{diff}}$,
after averaging $\pm k_{\textrm{eff}}$. The uncertainty per shot
is $\unit[39]{mrad}$, corresponding to $\sigma_{\textrm{ind}}=\unit[53]{mrad}$.\label{fig:supp-phi-diff}}
\end{figure}

\begin{figure}[H]
\begin{centering}
\includegraphics[bb=0bp 0bp 792bp 375bp,clip,width=0.9\textwidth]{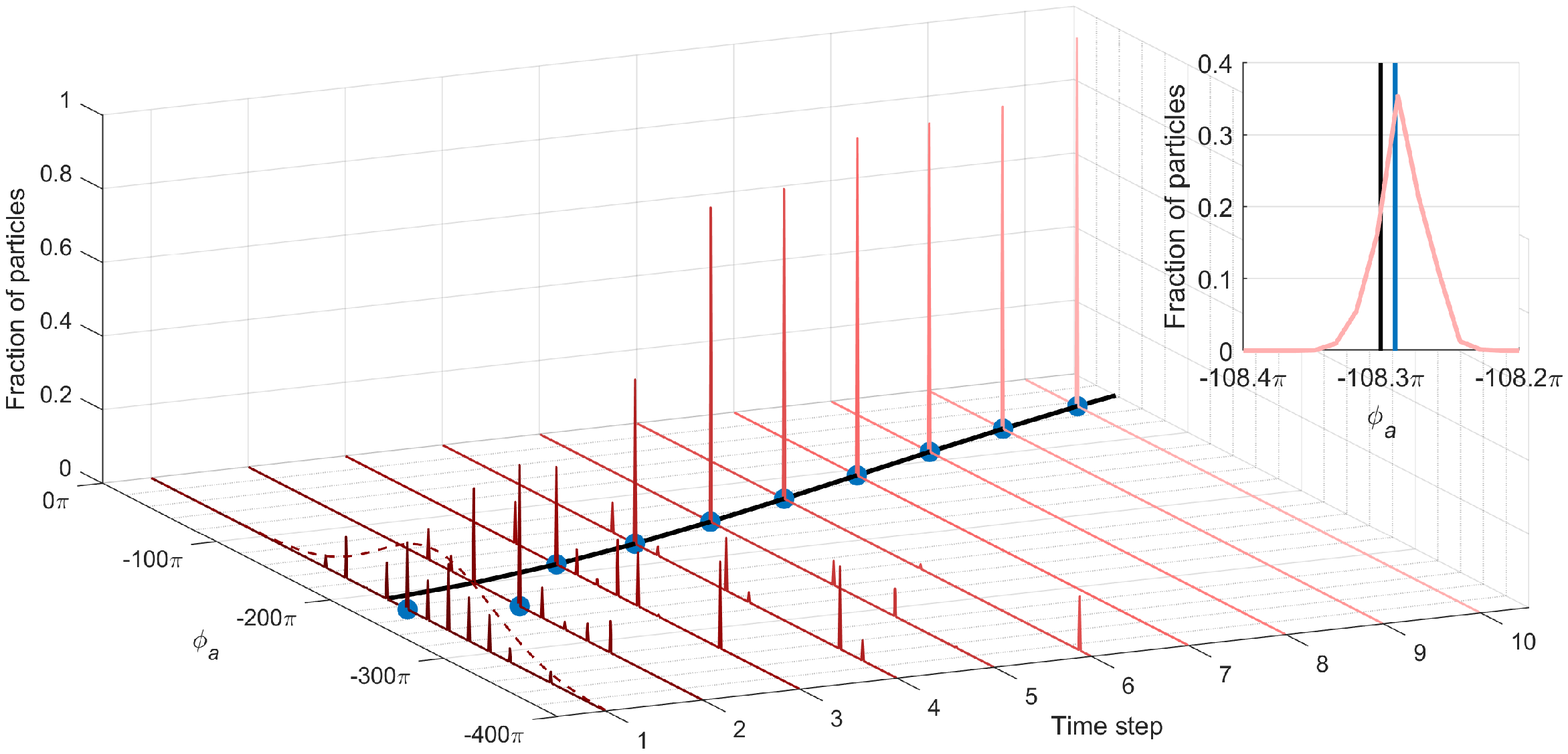}
\par\end{centering}
\centering{}\caption{Detailed presentation of the first ten time-steps of the particle
filter analysis for the dynamic signal presented in Fig.\,\ref{fig:particle}.
For each time step, we present a histogram of the $\phi_{a}$ values
of the particles, weighted based on their likelihood of describing
the actual measurement (red curves). Dashed curve in the first time
step is the initial distribution of particles, representing uncertainty
in the initial $\phi_{a}$ value. For reference, we present also the
input signal (black curve) and the $\phi_{a}$ values estimated from
ridge analysis of the particles data (blue markers). In the first
time steps, multiple solutions exist (spaced by multiples of $18\pi$
and $20\pi$ apart, for odd and even shots with $D=9$ and $10$,
respectively), and the ridge detection algorithm does not necessarily
select the correct one. After several time steps, most of the particles
(and eventually all) converge to follow the input signal. Inset: zoom-in
on the particle distribution in the 7th time step. \label{fig:supp-filter}}
\end{figure}

\newpage{}

\appendix
\onecolumngrid

\section*{Supplementary Information}

\setcounter{equation}{0}
\renewcommand{\theequation}{S\arabic{equation}}

\textbf{Raman beams generation and optics.} The Raman beams are derived
from a single distributed Bragg reflector laser diode tuned $\unit[1.7]{GHz}$
below the $F=2\rightarrow F'=1$ of the$\text{D}_{2}$ transition.
The Raman laser is phase modulated with an electro-optic modulator
(EOM) at $\unit[\sim6.834]{MHz}$, followed by dual-stage optical
amplification. The relatively small one-photon detuning causes significant
spontaneous scattering, and in fact limits our ability to implement
more than two interferometric sequences in a single shot due to loss
of contrast. The maximal in-fiber Raman beams intensity, including
all modulation sidebands, is $\unit[1]{W}$. For dual-$T$ operation,
we use only $\unit[0.5]{W}$ to reduce the velocity acceptance range
of the Raman pulses. The beams are collimated to a $\unit[40]{mm}$
diameter and have a circular polarization. After traversing the vacuum
chamber, they are retro-reflected from a mirror mounted on a piezo
tip-tilt stage and a passive vibration-isolation platform. Residual
vibrations are measured with a sensitive classical accelerometer and
the associated phase noise is subtracted in post-processing \citep{Merlet2009,Lautier2014}.

The Raman EOM is driven by a fixed $\unit[6.8]{GHz}$ microwave signal
mixed with a $\unit[\sim34]{MHz}$ signal from an agile direct digital
synthesizer (DDS). For dual-$T$ operation, each interferometer employs
one of two phase-coherent channels of the DDS, which are electronically
switched prior to each Raman pulse. The two signals are step-wise
chirped at an equal rate $\alpha$ with a relative offset of $\unit[80]{kHz}$
to address the two velocity classes. We set $\alpha=\alpha_{0}-k_{\text{eff}}\delta g$,
where $\alpha_{0}=$$k_{\text{eff}}g$ ($g$ is the local gravity
acceleration), and $\delta g$ is the simulated change in gravity.
$k$-reversal is achieved by flipping the sign of $\alpha$.

\textbf{Noise characterization.} We estimate $\sigma_{\phi,\textrm{est}}$
directly as the standard deviation of $\phi_{\textrm{est}}-\phi_{\textrm{in}}$.
Generally, we observe $\sigma_{\phi,\textrm{est}}$ in the range of
$\unit[155-180]{mrad}$ ($\unit[3.2-3.6]{\mu m/s^{2}}$) per shot,
with the dominant contribution attributed to residual vibration noise.
As such, it varies slightly between measurements done at different
times, depending on the environmental noise. We also observe some
dependence on $D$, with larger noise at lower $D$ values {[}Fig.\,\ref{fig:benchmark}(d){]},
which is in part attributed to weaker common-noise rejection within
each dual-$T$ shot.

The estimate of $\sigma_{\textrm{ind}}$ is based on the standard
deviation of $\phi_{\textrm{diff}}-\phi_{\textrm{diff,}0}$, with
$\phi_{\textrm{diff,}0}$ calculated from $\phi_{a}$ according to
Eq.\,(\ref{eq:rot_mat}). We note that for low $D$ values with negligible
outlier probabilities, $\sigma_{\textrm{ind}}$ can also be estimated
according to the standard deviation of $\delta\phi_{\textrm{ind,diff}}$,
which is the residual between measured $\phi_{\textrm{diff}}$ and
its nearest allowed discrete value, as defined in Eq.\,(\ref{eq:phi_diff_def}),
without using any prior information on $\phi_{a}$; however, when
the outlier fraction becomes significant, this method would result
in under-estimation of $\sigma_{\textrm{ind}}$. We observe $\sigma_{\textrm{ind}}$
in the range of $\unit[50-80]{mrad}$. We note that such values of
detection-induced noise would correspond in conventional atom interferometry
to detection signal-to-noise ratio, normalized by fringe contrast,
of 50 to 100 \citep{Yankelev2019}.
\end{document}